\documentstyle[12pt,epsf]{article}
\expandafter\ifx\csname amssym12.def\endcsname\relax \else\endinput\fi
\expandafter\edef\csname amssym12.def\endcsname{%
       \catcode`\noexpand\@=\the\catcode`\@\space}
\catcode`\@=11
\def\undefine#1{\let#1\undefined}
\def\newsymbol#1#2#3#4#5{\let\next@\relax
 \ifnum#2=\@ne\let\next@\msafam@\else
 \ifnum#2=\tw@\let\next@\msbfam@\fi\fi
 \mathchardef#1="#3\next@#4#5}
\def\mathhexbox@#1#2#3{\relax
 \ifmmode\mathpalette{}{\m@th\mathchar"#1#2#3}%
 \else\leavevmode\hbox{$\m@th\mathchar"#1#2#3$}\fi}
\def\hexnumber@#1{\ifcase#1 0\or 1\or 2\or 3\or 4\or 5\or 6\or 7\or 8\or
 9\or A\or B\or C\or D\or E\or F\fi}
\font\tenmsa=msam10 scaled\magstep1
\font\sevenmsa=msam7 scaled\magstep1
\font\fivemsa=msam5 scaled\magstep1
\newfam\msafam
\textfont\msafam=\tenmsa
\scriptfont\msafam=\sevenmsa
\scriptscriptfont\msafam=\fivemsa
\edef\msafam@{\hexnumber@\msafam}
\mathchardef\dabar@"0\msafam@39
\def\dashrightarrow{\mathrel{\dabar@\dabar@\mathchar"0\msafam@4B}}
\def\dashleftarrow{\mathrel{\mathchar"0\msafam@4C\dabar@\dabar@}}

\def\ulcorner{\delimiter"4\msafam@70\msafam@70 }
\def\urcorner{\delimiter"5\msafam@71\msafam@71 }
\def\llcorner{\delimiter"4\msafam@78\msafam@78 }
\def\lrcorner{\delimiter"5\msafam@79\msafam@79 }
\def\yen{{\mathhexbox@\msafam@55 }}
\def\checkmark{{\mathhexbox@\msafam@58 }}
\def\circledR{{\mathhexbox@\msafam@72 }}
\def\maltese{{\mathhexbox@\msafam@7A }}
\font\tenmsb=msbm10 scaled\magstep1
\font\sevenmsb=msbm7 scaled\magstep1
\font\fivemsb=msbm5 scaled\magstep1
\newfam\msbfam
\textfont\msbfam=\tenmsb
\scriptfont\msbfam=\sevenmsb
\scriptscriptfont\msbfam=\fivemsb
\edef\msbfam@{\hexnumber@\msbfam}

\def\widehat#1{\setbox\z@\hbox{$\m@th#1$}%
 \ifdim\wd\z@>\tw@ em\mathaccent"0\msbfam@5B{#1}%
 \else\mathaccent"0362{#1}\fi}
\def\widetilde#1{\setbox\z@\hbox{$\m@th#1$}%
 \ifdim\wd\z@>\tw@ em\mathaccent"0\msbfam@5D{#1}%
 \else\mathaccent"0365{#1}\fi}
\font\teneufm=eufm10 scaled\magstep1
\font\seveneufm=eufm7 scaled\magstep1
\font\fiveeufm=eufm5 scaled\magstep1
\newfam\eufmfam
\textfont\eufmfam=\teneufm
\scriptfont\eufmfam=\seveneufm
\scriptscriptfont\eufmfam=\fiveeufm

\csname amssym.def\endcsname
\newif{\ifcomentarios}
\comentariosfalse
\renewcommand{\theequation}{\thesection.\arabic{equation}}

\newcommand{\zerarcounters}
{
\setcounter{equation}{0}
\setcounter{theorem}{0}
}

\newcommand{\be}{\begin{equation}}
\newcommand{\ee}{\end{equation}}
\newcommand{\bma}{\begin{displaymath}}
\newcommand{\ema}{\end{displaymath}}
\newcommand{\bc}{\begin{center}}
\newcommand{\ec}{\end{center}}
\newcommand{\text}{\rm}

\newcommand{\uflex}
{{\scriptstyle {\raise 9pt\hbox{$\backslash$}\,\!\!\!\!\!\Bigg\vert}}}
\newcommand{\ncm}{\newcommand}

\ncm{\rncm}{\renewcommand}
\ncm{\id}{{\bf 1}}
\ncm{\beq}{\begin{equation}}
\ncm{\eeq}{\end{equation}}
\ncm{\ba}{\begin{array}}
\ncm{\bea}{\begin{eqnarray}}
\ncm{\beanon}{\begin{eqnarray*}}
\ncm{\ea}{\end{array}}
\ncm{\eea}{\end{eqnarray}}
\ncm{\eeanon}{\end{eqnarray*}}
\ncm{\fns}{\footnotesize}
\ncm{\setc}[1]{\setcounter{equation}{#1}}
\newcounter{eqnr}
\newenvironment{eqnarrayabc}{\stepcounter{equation}
  \setcounter{eqnr}{\value{equation}}\setc{0}
  \rncm{\theequation}{\thesection.\arabic{eqnr}\alph{equation}}
  \begin{eqnarray}}{\end{eqnarray}\setc{\value{eqnr}}}
\ncm{\eqboxabc}[3]{\newline\parbox[t]{1.5cm}{#1}\hfill
  \parbox[b]{12cm}{\begin{eqnarray*} #3\end{eqnarray*}}\hfill
   \parbox[b]{1.5cm}{\vspace{-0.0cm}\begin{eqnarrayabc}#2\end{eqnarrayabc}}\newline}
\ncm{\eqbox}[2]{\newline\parbox{1.5cm}{#1}\hfill
  \parbox{12cm}{\beanon #2\eeanon}\hfill
  \parbox{1cm}{\bea\eea}\newline}
\ncm{\nr}[1]{\parbox{1cm}{\begin{eqnarrayabc}#1\end{eqnarrayabc}}\\}

\ncm{\kal}[1]{\mbox{$\cal #1 $}}
\ncm{\mrk}[1]{\!\!\! #1 \!\!\!} 
\ncm{\qed}{\hspace*{0.4cm}\rule{0.24cm}{0.24cm}}  
\ncm{\mbold}[1]{\mbox{\boldmath $ #1 $}}   
\ncm{\bm}{\mbold}
\ncm{\str}{\stackrel}
\ncm{\sub}{\subset}
\ncm{\e}{\varepsilon}
\ncm{\ka}{\kappa}
\ncm{\inputc}[1]{\begin{center}\input{#1}\end{center}}
\ncm{\lto}{\longrightarrow}
\ncm{\x}{\times}
\ncm{\bmm}{\bm{\cal M}}
\ncm{\cp}{{\bf P}}    
\ncm{\bfp}{{\bf P}}
\ncm{\bmi}{\bm{i}}
\ncm{\bmom}{\bm{\om}}
\ncm{\bmOm}{\bm{\Om}}
\ncm{\res}{\restriction}
\ncm{\bmL}{\bm{\cal L}}
\ncm{\bmell}{\bm{\ell}}
\ncm{\bmE}{\bm{\cal E}}
\ncm{\bme}{\bm{e}}
\ncm{\bmpi}{\bm{\pi}}
\ncm{\bmr}{\bm{r}}
\ncm{\bmsigma}{\bm{\sigma}}
\ncm{\wt}{\widetilde}
\newcommand{\beaa}{\begin{eqnarray}}
\newcommand{\eeaa}{\end{eqnarray}}

\begin{document}

\author{{\bf Oscar Bolina}\thanks{Supported by FAPESP. Present
Address: Department of Mathematics, University of California at Davis, 
Davis CA 95616-8633, USA} \\ 
Instituto de F\'{\i}sica \\
Universidade de S\~ao Paulo \\
Caixa Postal 66318 \\
05315-970 S\~ao Paulo, Brazil \\
{\bf E-mail:} bolina@fma1.if.usp.br \\
\and {\bf J. Rodrigo 
Parreira}\thanks{Supported by CNPq.} \\
Department of Physics \\
Princeton University \\
P.O. Box 708 \\
08544-0708 Princeton NJ USA\\
{\bf e-mail:} parreira@math.princeton.edu \\
}
\title{\vspace{-1in}
{\bf A Cluster Expansion for Dipole Gases}}
\date{}
\maketitle
\begin{abstract}
\noindent
We give a new proof of the well-known upper bound on the correlation
function of a gas of non-overlapping dipoles of fixed length and 
discrete orientation working directly in the charge representation,
instead of the more usual sine-Gordon representation.

\noindent
{\bf Key words:} Dipole Gas, Charge Representation,
Cluster Expansion, Correlation Function \hfill \break
\end{abstract}


\section{Introduction}
\zerarcounters

Although dipole gases themselves do not exhibit the Kosterlitz-Thouless
transition, they have been used as a natural starting point for the study
of this transition in Coulomb systems at low temperatures and small
activities \cite{FS1}.
\newline
Differently from the behavior of correlations in Coulomb systems,
charge correlations in dipole gases have lower and upper bounds with
power law decay. The lower bound is a consequence of Jensen's inequality
in the charge variables \cite{FS1}. The main ingredient to obtain an upper
bound is the {\it sine}-Gordon representation of the gas. Using this
representation Frohlich and Spencer \cite{FS2} have proved a theorem:
\newline
{\bf Theorem} {\it The two-point correlation function of external
charges for dipoles with hard core is bounded above by}
\beq
G(x,y) \leq C_{\beta} e^{-m \log|x-y|} \label{1}
\eeq
{\it with} $C_{\beta} < \infty$ {\it and} 
\[
m=\frac{\beta}{2 \pi}(1+(1+\epsilon)\beta z e^{-{\beta}/{16}})^{-1}
\]
{\it for any} $\epsilon >0$, {\it where} $\beta$ {\it is the inverse
temperature and z is the activity}. 
\newline
The correlation function above can be equivalently written as 
$G(x,y)=e^{-\beta V_{eff}(x,y)}$, where the effective potential
$V_{eff}(x,y)$ is the difference between the free energy of the
system in the presence of the external charges and the free energy 
in the absence of the external charges. The physical content of the
above theorem is that the long range dependence on the position of the
external charges in (\ref{1}) implies that dipoles do not screen
\cite{GJ}.
\newline 
In this paper we dispense with the usual {\it sine}-Gordon
representation and establish the upper bound (\ref{1}) for a gas of
non-overlapping dipoles of fixed length and discrete orientation,
working directly in the charge (or gas) representation.
\newline
To obtain the above result we use an electrostatic inequality in
the form of charge translations in order to drive down the activity of
the dipoles and obtain estimates for small activities, followed by a
direct cluster expansion of the partition function to extract the
fall-off of the correlation. (The extension of the technique to 
Coulomb gases follows along the same lines). 
\newline
Our method provides an alternative procedure for dealing with systems 
for which no {\it sine}-Gordon representation is available. (See 
\cite{MK} for the use of a charge representation in connection with
percolation models.)  In particular, we hope that the charge
representation will enable us to give a direct proof of the analyticity
of the pressure both for dipole and Coulomb gases.


\section{Dipole Gases}
\zerarcounters

We consider a lattice dipole gas with discrete orientation in a
finite region $\Lambda$. Let $L=4r{\bf Z}^{2}$ be a sublattice of
$\Lambda$ of mesh 4r, contained in a box $\hat\Lambda$ inside
$\Lambda$ and concentric to it. A configuration of a lattice dipole is
described by a function $\vec{\mu}_{L}=\{ \vec{\mu}(j) \}$, where
$\vec{\mu}(j)=(\mu_{1}(j),\mu_{2}(j))$ is the total dipole moment at the
site $j \in L$. The dipoles are non-overlapping if their fixed length is
{\it r}. For simplicity we shall take $r=1$.
\newline
The partition function is given by 
\begin{equation}
{\cal Z} =\sum_{ \vec{\mu}_{L}= \{\vec{\mu}(j),  {j \in L} \} }
e^{-\frac{\beta}{2} E(\vec{\mu}_{L})} \prod_{j \in L}
d\lambda(\vec{\mu}(j)), \label{2}
\end{equation}
where the interaction energy is
\[
E(\vec{\mu}_{L})= \sum_{j,k} \sum_{\alpha, \gamma}
\mu_{\alpha}(j)\mu_{\gamma}(k) W_{\alpha \gamma}(j,k)
\]
and the two-body potential is given by
\[
W_{\alpha \gamma}(j,k)=\partial_{\alpha} \partial_{\gamma} C(j,k) 
\]
where $\partial_{\alpha}$ is the usual finite difference derivative in
the $\alpha$-direction, and $C(j,k)$ is the Green's function of a
lattice Coulomb potential in two dimensions.
\newline
The "a priori" distribution $d\lambda$ of $\vec{\mu}(j)$ is given by
the measure
\[
d\lambda(\vec{\mu})=\delta(\vec{\mu})+z\delta(|\vec{\mu}|^{2}-1)
\]
where {\it z} is the activity, the same for all the dipoles.
\newline
The correlation function is defined by
\[
G(x,y)=\frac{{\cal Z}^{\xi}_{L}(x,y)}{{\cal Z}_{L}},
\]
where the ${\cal Z}^{\xi}_{L}(x,y)$ is the partition function in the
presence of the external charges, given by
\[
{\cal Z}^{\xi}(x,y) =\sum_{ \vec{\mu}_{L} = 
\{ \vec{\mu}(j), {j \in L} \} } e^{-\frac{\beta}{2} 
E(\vec{\mu}_{L}+{\xi}(j))}
\prod_{j \in L} \lambda(\vec{\mu}(j)).
\]
where ${\xi}(j)=\xi(\delta_{j,x}-\delta_{j,y})$, and 
${\cal Z}_{L}={\cal Z}^{\xi=0}_{L}$.
\newline
Throughout this paper we suppose that the external charges are located
far apart from each other, and that {\it x} and {\it y} do not belong
to $\hat{\Lambda}$. 


\section{Electrostatic Inequality}
\zerarcounters

The first step to obtain an upper bound on the correlation function
consists in driving down the activity of the dipoles, so as to obtain 
estimates valid for small values of {\it z}. This is done by the
electrostatic inequality. To apply this inequality, it is more convenient
to define a dipole charge density by
\[
\rho(v) = \sum_{j,\alpha} \mu_{\alpha}(j)(\delta_{j,v}-
\delta_{j,v+e_{\alpha}})
\]
and write the electrostatic energy as $E(\vec{\mu}_{L})=E(\rho)= (\rho,
-\Delta^{-1} \rho )$.
If we carry out the summation on the dipole moment variables in (\ref{2}),
we get
\[
{\cal Z}=\sum_{\rho}
e^{-\frac{\beta}{2} E(\rho)} \prod_{j,\alpha} z_{\alpha}(j)
\]
where $z_{\alpha}(j)=z(j)^{|\mu_{\alpha}(j)|}$. Because of the discrete
orientation of the dipoles, $\mu_{\alpha}(j)$ takes the values $\{0, \pm
1 \}$ in such a way that $\mu_{1}(j)\mu_{2}(j)=0$. 
\newline
Now we replace the charge density $\rho$, whose  support lies 
in $\hat\Lambda$, by a charge density $\bar{\rho}$, whose support lies in
$\Lambda$, in such a way that the interaction energy between the two
renormalized densities remain unchanged, but that the self-energy of
$\bar{\rho}$ is smaller than the self-energy of $\rho$. We use the
difference in self-energies to renormalize the activities \cite{FS1}.
\newline
The renormalized charge density is given by 
\beq
\bar{\rho}=\rho + \frac{\Delta \rho}{||\Delta||}. \label{3}
\eeq
For the purpose of renormalizing the activity of the dipoles, we write
\beq
{\cal Z}=\sum_{\bar{\rho}} e^{-\frac{\beta}{2} E(\bar{\rho})}
e^{-\frac{\beta}{2}[E(\rho)-E(\bar{\rho})]}
\prod_{j,\alpha} z_{\alpha}(j). \label{4}
\eeq
The difference in self-energy is estimated as follows: 
\[
E(\rho)-E(\bar{\rho}) = \frac{2}{||\Delta||} (\rho,\rho)
- \frac{1}{{||\Delta||}^{2}} (\rho, -\Delta \rho) 
\geq \frac{1}{||\Delta||} (\rho,\rho).
\]
So that we obtain
\beq
{\cal Z}=\sum_{\bar{\rho}} e^{-\frac{\beta}{2} E(\bar{\rho})}
\prod_{j,\alpha} {\bar{z}}_{\alpha}(j), \label{5}
\eeq
with the estimate ${\bar{z}}_{\alpha}(j) \leq z_{\alpha}(j) 
\exp\{-({\beta}/{|| \Delta ||}) \rho_{\alpha}^{2}(j)\}$.
\newline
Since the activities are all equal, all the renormalized activities can be
bounded by the same constant (take $|| \Delta || = 8$) :
\beq
{\bar{z}}_{\alpha}(j) \leq \bar{z}=z \exp\{-\frac{\beta}{16}\}, \label{6}
\eeq
which can be made arbitrarily small for $\beta$ sufficiently large.


\section{The External Charges}
\zerarcounters

We apply the above reasoning to the partition function of the system 
in the presence of external charges, and obtain 
\[
{\cal Z}^{\xi}(x,y)=\sum_{\bar{\rho}}
e^{-\frac{\beta}{2} E(\bar{\rho}+\xi)}\prod_{j,\alpha}
\bar{z}_{\alpha}(j).
\]
However, in this case we will need to renormalize the external
charges as well. This is done by the shift 
\[
\sigma=\bar{\rho}+ \xi + \frac{\gamma}{\beta}(\Delta a)(j)
\] 
where $\gamma$ depends on $\beta$ in a way to be specified later, and
\beq
a(j)=C(j,x)-C(j,y). \label{7}
\eeq
The function $a(j)$ satisfies, for large $|x-y|$ and for $|j-j'|=1$, 
the bounds [3, chapter 7]:
\beq
|a(j)-a(j')| \leq {\rm const} \left( \frac{1}{|j-x|+1} + 
\frac{1}{|j-y|+1} \right), \label{8}
\eeq
\beq
|a(j)-a(j')| \leq {\rm const} \frac{|x-y|}{|j|^{2}}, \label{9}
\eeq
From the definition of the Laplacian operator and (\ref{7}), we also have 
\beq
\sum_{|j-j'|=1} [a(j)-a(j')]^{2} \leq (a, -\Delta a) = a(x)-a(y)
\approx \frac{1}{\pi} \log |x-y|. \label{10}
\eeq
After the shift we get, following Eqs. (\ref{4}) and (\ref{5}), 
\beq
{\cal Z}^{\xi}(x,y)=e^{(\frac{\gamma^{2}}{2 \beta}-\gamma)(a, -\Delta a)}
Z_{1}, \label{11}
\eeq
where 
\beq
Z_{1}=\sum_{\sigma} e^{-\frac{\beta}{2} E(\sigma)}
\prod_{j,\alpha} \bar{z}_{\alpha}(j) e^{\gamma_{\alpha}(j)}, 
\label{12}
\eeq
with 
\beq
\gamma_{\alpha}(j)=\gamma \mu_{\alpha}(j)
\delta_{\alpha}\bar{a}(j), \label{13}
\eeq
and the definition of $\bar{a}(j)$ follows from (\ref{3}). 


\section{The Cluster Expansion}
\zerarcounters

To derive an upper bound on the partition function (\ref{12})
we write
\[
\prod_{j,\alpha} \bar{z}_{\alpha}(j) 
e^{\gamma_{\alpha}(j)}= \prod_{j,\alpha}
\bar{z}_{\alpha}(j) \cosh{\gamma_{\alpha}(j)} + R
\]
where ($\mu=1,2$)
\beq
R=\prod_{j,\alpha}(\bar{z}_{\alpha}(j) 
\cosh{\gamma_{\alpha}(j)}+\bar{z}_{\alpha}(j) 
\sinh{\gamma_{\alpha}(j)})- \prod_{j,\alpha}
\bar{z}_{\alpha}(j) \cosh{\gamma_{\alpha}(j)} \label{R}
\eeq
Thus (\ref{12}) becomes
\beq
Z_{1}=\sum_{\sigma} e^{-\frac{\beta}{2} E(\sigma)}  
\left \{ \prod_{j,\alpha} \bar{z}_{\alpha}(j) 
\cosh{\gamma_{\alpha}(j)} +  R \right \}  \label{Z1}
\eeq
The function $\gamma_{\alpha}(j)$ depends on $\delta_{\alpha}\bar{a}(j)$.
However, note that 
\[ 
\delta_{\alpha}\bar{a}(j) = \bar{a}(j)-a(j+e)=a(j)-a(j+e) \;\;\;
{\rm when} \;\;\; j \neq x,y, 
\]
since $(\Delta a)(j)=0  \;\;\; {\rm for} \;\;\; j \neq x,y$.
\newline
From the definition (\ref{13}), and the bounds (\ref{8}), (\ref{9}) we
see that the argument of $\cosh(\cdot)$ is small for $|j-x| \gg \gamma$
and $|j-y| \gg \gamma$. Thus we can write 
\[
\cosh{\gamma_{\alpha}(j)}-1 \leq 
\frac{1}{2}(1+\epsilon) {\gamma^{2}[a(j)-a(j')]}^{2}
\]
for some $\epsilon >0$.
\newline
For $|j-x| \leq O(\gamma), |j-y| \leq O(\gamma)$, we estimate
$\cosh(\cdot)-1$ by a constant.
\newline
In both cases we can take {\it z} sufficiently small so that
the (\ref{R}) term can be bounded by a constant for large $\beta$.
\newline
Under these circumstances, the {\it R}-terms in 
(\ref{Z1}) can be neglected: 
\beq
Z_{1} \leq K_{\beta} \sum_{\sigma}
e^{-\frac{\beta}{2} E(\sigma)} \prod_{j,\alpha}
\bar{z}_{\alpha}(j) \cosh{\gamma_{\alpha}(j)} \label{MM}
\eeq
for some positive constant $K_{\beta}$.
\newline
Now we develop a cluster expansion for $\prod_{j,\alpha}
\cosh{\gamma_{\alpha}(j)}$ as follows  
\beq
\prod_{j,\alpha} [1+(\cosh{\gamma_{\alpha}(j)}-1)]=
1 + S_{1}+S_{2}+S_{1}S_{2} \label{FF} 
\eeq
where
\beq
S_{\alpha} = \sum_{m_{\alpha} \geq 1} \sum_{j_{1},...,j_{m_{\alpha}}} 
D_{\alpha}(j_{1}) . \;\;  .  \;\; . D_{\alpha}(j_{m_{\alpha}}),
\label{CLU}
\eeq
with
\beq
D_{\alpha}(j_{i})=\cosh{\gamma_{\alpha}(j_{i})}-1.
\eeq
The sum in (\ref{CLU}) above runs over all subsequences of  
$\{j_{1},...j_{m_{\alpha}} \}$, $m \geq 1$. 
\newline
We can distribute the activities inside $S_{\alpha}$, to get
\beq
Z_{1} \leq K_{\beta} \left \{ {\cal Z} + \sum_{\sigma}
e^{-\frac{\beta}{2} E(\sigma)} (s_{1}+s_{2}+s_{1}s_{2}) \right \}, 
\label{19}  
\eeq
where $s=s_{1}+s_{2}+s_{1}s_{2}$ and, as in (\ref{CLU}), 
\[
s_{\alpha} = \sum_{m_{\alpha} \geq 1} \sum_{j_{1},...,j_{m_{\alpha}}}
\bar{z}_{\alpha}(j_{1})D_{\alpha}(j_{1}) . \;\;  .  \;\; . 
\bar{z}_{\alpha}(j_{m_{\alpha}})D_{\alpha}(j_{m_{\alpha}}).
\]
Thus we see that $Z_{1}$ admits the upper bound
\beq
Z_{1} \leq K_{\beta} {\cal Z}(1 + s_{1}+s_{2}+s_{1}s_{2}).
\eeq
Here the {\it s} terms in (\ref{19}) have been taken with
$\mu_{\alpha}(j)=1$ for all {\it j}, $\alpha$, and we have used that
${\cal Z} >1$.
\newline
If we {\it undo} the cluster expansion (\ref{FF}) we obtain
\beq
Z_{1} \leq K_{\beta} {\cal Z} \prod_{j,\alpha} (1+\bar{z}_{\alpha}(j)
D_{\alpha}(j)), \label{UN}
\eeq
with $\mu_{\alpha}(j)=1$ for all {\it j}, $\alpha$.
\newline
Substituting (\ref{UN}) in (\ref{12}) and using (\ref{6}) and (\ref{10}),
we get
\beq
{\cal Z}^{\xi}(x,y) \leq  K_{\beta} \, {\cal Z} \,  
e^{ \{ \frac{\gamma^{2}}{2\beta} + 
\frac{1}{2}(1+\epsilon)\bar{z}\gamma^{2} - \gamma \} 
(a, -\Delta a)}. 
\eeq
The optimal choice for $\gamma$ is 
$\gamma=\beta(1+(1+\epsilon)\bar{z}\beta))^{-1}$, for which we finally
get
\[
G(x,y) \leq K_{\beta} e^{-\frac{\beta}{2 \pi}
\frac{1}{1+(1+\epsilon)\bar{z}\beta} \log{|x-y|}},
\]
which concludes the argument.
\vskip .2 cm
\noindent
{\small {\it Acknowledgement.}} {\footnotesize The authors are very 
grateful to Dr. D.U.H.Marchetti for help in understanding some of the
estimates.} 


\end{document}